\documentclass[12pt,aasms4]{aastex}

\begin{document}
\title{\bf Phase-Resolved Infrared {\it H}- and {\it K}-band Spectroscopy of 
EF Eridani\footnote{Partially based on observations obtained at La Silla (ESO proposal
69.D-0142(A))}}

\author{Thomas E. Harrison\footnote{Visiting Astronomer, Kitt Peak National 
Observatory, National
Optical Astronomy Observatory, which is operated by the Association of 
Universities for Research in Astronomy, Inc., under cooperative agreement
with the National Science Foundation.}}
\affil{Astronomy Department, New Mexico State University, Las Cruces, 
New Mexico, 88003}
\authoremail{tharriso@nmsu.edu}

\author{Steve B. Howell\footnote{Visiting Astronomer, NTT Telescope at La Silla
Operated by ESO}}
\affil{WIYN Observatory and National Optical Astronomy Observatory, 950
North Cherry Avenue, Tucson, AZ 85719}
\authoremail{howell@noao.edu}

\author{Paula Szkody}
\affil{Department of Astronomy, University of Washington, Box 351580, Seattle,
WA 98195}
\authoremail{szkody@alicar.astro.washington.edu}

\author{Derek Homeier}
\affil{Department of Physics and Astronomy, University of Georgia, Athens,
GA 30602-2451}
\authoremail{derek@physast.uga.edu}

\author{Joni J. Johnson, Heather L. Osborne}
\affil{Astronomy Department, New Mexico State University, Las Cruces, 
New Mexico, 88003}
\authoremail{jojohnso@nmsu.edu, hosborne@nmsu.edu}

\begin{abstract}
We present new phase-resolved {\it H} and {\it K}-band spectroscopy of the 
ultra-short period magnetic cataclysmic variable EF Eri in its current,
prolonged ``low'' state obtained using 
NIRI on Gemini-North, and NIRSPEC on Keck II. These new data show that the 
{\it H}-band spectrum of EF Eri appears to be dominated by cyclotron emission 
during the entire orbital cycle. The {\it K}-band spectrum of EF Eri is 
likewise dominated by cyclotron emission during most of an orbital period, but 
near binary phase
0.0, the secondary star spectrum may be visible. The lack of
strong CO or CH$_{\rm 4}$ absoprtion features, and the weakness of the
water vapor features in this spectrum, however, suggests the possibility of 
peculiar abundances for Carbon and/or Oxygen. We have used the \texttt{PHOENIX} 
stellar atmosphere code to produce
model brown dwarf spectra with non-solar abundances of Carbon, Nitrogen
and Oxygen and achieved limited success in fitting the observed spectra.
We conclude that strong, and highly variable cyclotron emission is responsible for the
photometric variation previously reported for EF Eri.
The nature of this cyclotron emission is complex: the {\it H}-band spectra
show that the dominant cyclotron harmonic at phase 0.5 peaks
at 1.65 $\mu$m, but at phase 0.0, the harmonic peaks near 1.72 $\mu$m.
At phase 0.5, there is another cyclotron feature present that peaks in between
the {\it H} and {\it K} bands (near 1.93 $\mu$m), but at phase 0.0, no 
such feature is present. These data suggest that cyclotron emission from both
poles is occurring. In the high state, the cyclotron emission
has been modeled as coming from the pole that is oriented towards the
secondary star. One interpretation for the phase 0.5 cyclotron emission is
that it originates from the opposite pole. In its current ultra-low
state, EF Eri reveals no outward signs of accretion (such as H I emission),
but continues to have a few, strong cyclotron features. Thus, EF Eri
joins the small group of magnetic cataclysmic variables whose accretion rate 
is so low that they are in the ``bombardment scenario'' regime. 
\end{abstract}


\begin{flushleft}
\end{flushleft}

\section{Introduction}
Polars are magnetic cataclysmic variables that have white dwarf primaries with 
strong magnetic fields that dramatically effect the accretion flow 
from their cool Roche lobe filling companions. Instead of forming a disk, the 
accreted matter flows in a stream from near the L$_{\rm 1}$ point to 
one, or both, of the magnetic poles of the white dwarf primary 
(see Wickramasinghe \& Ferrario 2000 for a complete review). For reasons not 
yet identified, the normal ``high state'' mass transfer rate ($\approx$ 10$^{-11}$ M$_{\sun}$ yr$^{-1}$) ceases, and the system transitions to a ``low state". 
For the most extreme examples, accretion rates as 
low as 10$^{-13}$ M$_{\sun}$ yr$^{ -1}$, or lower, have been inferred (Reimers 
et al. 1999, Schwope et al. 1999, Szkody et al. 2003). When in these low 
states, polars should present an excellent opportunity to examine the very low
mass secondary stars of the shortest period cataclysmic variables (CVs). 

EF Eri is an ultra-short period (P$_{\rm orb}$ = 81 min) polar that entered a 
low state in 1997. Harrison et al. (2003) presented simultaneous 
multi-wavelength optical and infrared photometry for EF Eri in its low state in 
which they believed they had detected the irradiated brown dwarf-like
secondary star of this system. Their data showed that in the {\it BVRIJ} 
bandpasses, the light curves were dominated by emission 
from the white dwarf and some residual cyclotron emission, while in the 
{\it H}- and {\it K}-bands, the light curves exhibited large amplitude 
($\Delta$K = 0.8 mag) variations that repeated once per orbital cycle. 
Harrison et al.  
were able to model these multi-wavelength data with 
an extremely cool (T$_{\rm eff} \leq$ 1,000 K) secondary star that is 
being irradiated by the primary. In this model, the irradiated hemisphere
of the secondary star is heated to T$_{\rm eff} \approx$ 1,500 K. If this 
model were correct, then the spectroscopic appearance of the secondary star 
should show dramatic changes over an orbital cycle, oscillating between a 
state where CO absorption is strong, as in the L-dwarfs, to one where methane 
might dominate, as seen in T-dwarfs. We have obtained new phase-resolved 
{\it H-} and {\it K}-band spectra of EF Eri using NIRSPEC on KECK II, and 
{\it K}-band spectra using NIRI on Gemini-North, to test the irradiated
brown dwarf model.

\section{Observations}

EF Eri was observed continuously from 7:21 until 9:00 UT on 24 December
2002 using NIRI\footnote{http://www.us-gemini.noao.edu/sciops/instruments/niri/NIRIIndex.html}
 on Gemini-North in queue scheduled mode. For these observations
the F/6 camera, a four pixel slit and the {\it K-grism} were used to produce 
{\it K-}band spectra with a resolution of R $\approx$ 780. Individual spectra, 
with exposure times of 120 s, were obtained at five different positions along 
the slit. After each exposure, the telescope was offset by 6" along the slit,
and a new exposure started. In this way forty two individual spectra were
obtained in 100 minutes, covering 123\% of an orbital cycle for EF Eri.
In addition to this set of continuous spectra, four additional spectra
were obtained in the period between 6:10 and 6:19 UT.

To remove telluric features in the spectra of EF Eri, the G2V star HD16358 was 
observed using an identical procedure, but with exposure times of 5 s. At the 
time of the observation, HD16358 had an airmass of 1.44, which is very close 
to the mean airmass attained by EF Eri during the 100 minute observation period.
The resulting dataset was reduced using a procedure common for reducing
infrared spectral data in which the images are grouped into sets of five 
individual exposures. A median of four of these exposures is subtracted 
from the spectrum of interest to remove the sky contamination. A spectrum is 
then extracted from this sky-subtracted image using the normal longslit 
spectroscopic reduction packages within IRAF, and the process is repeated. 
Each individual spectrum was wavelength calibrated using the spectrum of an 
argon lamp obtained shortly before our observations began.
The wavelength calibrated spectra of EF Eri were then divided by the 
median of the five wavelength calibrated spectra of HD16358 to remove 
telluric absorption features. The latter procedure introduces spurious 
features into the spectra of interest due to weak atomic and molecular
absorption lines in the G2V spectrum. To remove these weak features, we have 
employed the procedure described by Maiolino, Rieke, \& Rieke (1996).

With only a single standard to work with, the telluric 
correction for the entire data set is not as good as can be produced 
(see Harrison et al.  2004) when a large number of such spectra are available.
This is evidenced by dips at 2.0 $\mu$m
and 2.05 $\mu$m in the NIRI spectra presented below.
The flux calibration of these data was fairly good, and our spectra revealed
the same large amplitude ($\Delta${\it K} = 0.8 mag) flux variations
as reported by Harrison et al. (2003). We have assumed that EF Eri has
not changed its photometric behavior between the time of the last photometric 
data set 
(see below) and the epoch of these observations, allowing us to use the 
{\it K}-band light curve to produce flux-calibrated spectra.

EF Eri was observed using NIRSPEC\footnote{http://www2.keck.hawaii.edu/inst/nirspec/nirspec.html} on Keck II on 6 September 2003. We used NIRSPEC in the
low-resolution mode with the 0.38'' slit. We employed the four-nod script
to obtain spectra at four different positions along the slit.
 For the {\it H}-band spectra the
grating tilt was set to cover the wavelength range 1.54 to 1.83 $\mu$m with
a dispersion of 2.91 \AA/pix. The {\it K}-band setup covered the 
wavelength range 2.00 to 2.42 $\mu$m, with a dispersion of 4.27 \AA/pix.
Due to the strong curvature of the spectra produced by NIRSPEC,
each spectrum has a slightly different wavelength coverage
that depends on the position of the object along the slit.
This effect will be apparent in some of the NIRSPEC spectral sequences presented
below.  The {\it H}-band data were obtained from 12:29 to 13:57 UT, covering 
109\% of an orbital cycle. The {\it K}-band spectra were obtained in the 
interval 14:24 to 15:25 UT, and covered 88\% of an orbital period. HD16358
was used as the telluric standard for both bandpasses. While the
conditions were photometric, we found that the nodding routine did not
reliably place our targets exactly on the slit during the four-position
nod, and thus we were unable to extract photometric information from these data.
We have decided to use the infrared light curves to produce 
flux-calibrated spectra.

Three months before the Gemini observations, on 2002 September 24, we were 
able to obtain additional, simultaneous {\it JHK} photometry of EF Eri using
SQIID on the KPNO 2.1 m. These data were reduced in the same fashion as our
earlier observations
(see Harrison et al. 2003 for full details). The conditions during this
run were poorer than those of our 2001 run, and
the {\it J}-band light curve (see Fig. 1) has a lower S/N than the earlier
effort. The {\it H}- and {\it K}-band data, however, reveal identical light 
curves to the 2001 data set, showing no evidence for changes in either the 
absolute flux levels, or in the amplitude of the variations.

Harrison et al. (2003) found that the phasing of EF Eri by Bailey et al. 
(1982), where phase 0 corresponded to the appearance of a sharp
dip in the {\it J}-band light curve, appeared to be identical with binary 
orbital phase. We have used the same set of ephemerides to phase the data here.
Because of EF Eri's extremely short period, the spectral smearing caused
by the orbital motion is significant. We have used the evolution
code predictions by Howell, Nelson, \& Rappaport (2001) to estimate the
masses of the two components in the EF Eri system as M$_{\rm 1}$ = 0.6
M$_{\sun}$ and M$_{\rm 2}$ = 0.06 M$_{\sun}$. Using an orbital inclination
of 45$^{\circ}$ (see Harrison et al. 2003, and references therein), we
estimate the radial velocity semi-amplitude of the secondary in EF Eri
to be K$_{\rm 2}$ = 340 km s$^{\rm -1}$. The full orbital variation corresponds 
to only about two resolution elements in the NIRI spectra, but six 
resolution elements for the Keck spectra. Thus,
to allow us to coadd the individual spectra to search for features
from the secondary star, we have Doppler corrected our data using this 
estimated value for K$_{\rm 2}$.

In our ongoing program of continued, occasional monitoring EF Eri, we have 
obtained a new optical
spectrum using the 
EMMI\footnote{http://www.ls.eso.org/lasilla/Telescopes/NEWNTT/emmi/} 
spectrograph in the Red Imaging Low Dispersion (RILD) mode on the {\it NTT}. 
This spectrum, shown in Figure 2, was obtained
on 10 August 2002, is a 600 s exposure, and has a resolution of 2.3 \AA. It is 
interesting
to compare this spectrum with those presented in Harrison et al. (2003),
and Beuermann et al. (2000): the Zeeman absorption features are now more
pronounced, and the emission from H I has continued to decline to the point
where it is no longer detectable.

\section{Results}

In the preceding section we noted that we have produced flux-calibrated
spectra by using the infrared light curves for EF Eri. We did this so as 
to produce a consistent set of data that would enable us to ignore
slit losses in either set of spectra. Comparison of data sets obtained on widely
separated epochs like those analyzed here could be compromised by intrinsic
variations in EF Eri. While we are confident about the photometric ``stability''
of EF Eri up to, and including the Gemini data, there is no assurance that
EF Eri was in the same state when we observed it with Keck. But to make
sense of the Gemini NIRI data has required us to assume that the 
{\it H}-band spectra obtained with NIRSPEC represents a similar state in
EF Eri. The fact that the NIRSPEC {\it K}-band data had similar phase-dependent
morphology gives us confidence in this assumption.

In Fig. 3 we present the {\it H}- and {\it K}-band spectra for EF Eri at
a number of orbital phases. Each of these spectra is the median of four 
({\it H}) or five ({\it K}) individual spectra. The {\it H}-band spectra in this plot have
been produced as described above. The telluric correction scheme of
Maiolino et al. (1996) truncates the blue end of the {\it K}-band at
1.94 $\mu$m. Due to the blueward rise seen in the NIRI spectra of EF Eri, we
have also reduced the {\it K}-band spectra by simple raw division of these 
data by the telluric standard, and then multiplied the divided spectra
by the spectrum of the appropriate blackbody. This allowed us to extend our 
wavelength coverage 
down to $\approx$ 1.92 $\mu$m. This latter method, however, introduces weak 
``emission'' features from the G-star division (see Fig.1 in Harrison et al. 
2004).  Given the low S/N of our data set, the only obvious such feature is 
the false H I Br-$\gamma$ emission line at 2.16 $\mu$m visible in the
spectra with orbital phases near 0.5. 

\subsection{The {\it H}-band Spectra}

During one orbital cycle, the {\it H}-band spectra undergo remarkable
changes. Near phase 0.0 (i.e., minimum light), the spectra show a pronounced
blueward rise. Note that the data have relatively high S/N in this region,
as it is not near the blue cutoff of the {\it H}-band. Alternatively, one could
suggest that there is a single, broad absorption feature located at
1.574 $\mu$m. After this dip, the spectrum rises to a broad maximum whose
peak is near 1.72 $\mu$m. As we show below, this is not the expected behavior 
for the spectrum of a normal late-type/brown dwarf star.

As the orbital phase progresses from phase 0.0, the blueward rise rapidly
disappears, and the flux peak in the spectrum moves to the blue. It is
unclear whether this blue feature completely disappears, or is simply
hidden by the much higher flux levels associated with the phase 0.5 cyclotron 
emission.  During the transition to maximum light, a weak rise at the extreme 
red end of the {\it H}-band appears. 
While the S/N of the data in this region is quite low, the combination of the
{\it H} and {\it K}-band data sets suggest that this rise is real, and is due 
to a broad feature that peaks somewhere between the edges of the {\it H} and 
{\it K} bandpasses, in the middle of the region strongly affected by telluric
water vapor absorption. As the orbital phase continues to increase on its
way back to phase 0.0, the spectra retrace their pre-maximum evolution.

\subsection{The {\it K}-band Spectra}

The {\it K}-band spectra obtained with NIRI are shown
in Fig. 3, and the {\it K}-band spectra obtained with NIRSPEC are presented in 
Fig. 4. Both data sets show similar evolution over an orbital cycle,
suggesting that little has changed in the intervening nine months. Near phase 
0.0, the spectrum shows a continuum that is 
consistent with that of a late-type star, with declining fluxes at both
blue and red ends, presumably caused by water vapor absorption in the
photosphere of the secondary star. The co-addition of the NIRSPEC
phase 0.0 spectra (Fig. 5) do not confirm this, however, with no clear sign
of the expected spectral features for a late-type star. {\it This is
true even when we correct the spectra for the predicted radial velocity
of the secondary star.} While
a feature consistent with $^{\rm 12}$CO$_{\rm (2,0)}$ (at 2.294 $\mu$m) 
appears to be present, other features with a similar depth do not correspond 
to any of the expected atomic features (e.g., Na I at 2.20 $\mu$m) for
a cool star. By coadding
the entire NIRSPEC data set, the first overtone feature of $^{\rm 12}$CO$_{\rm (2,0)}$ appears to remain marginally detectable. This would be the first 
detection of the secondary star in EF Eri.

As the orbital phase advances, the spectra show rapid changes
 with the development of a pronounced hump blueward of 2.10 $\mu$m, and 
an increasingly steeper
continuum beyond 2.2 $\mu$m. This blue hump is better defined in the NIRI
spectra due to their extended blue coverage. Note that the telluric water
vapor absorption below 2.0 $\mu$m is very strong, and thus the shape of the 
continuum here is highly sensitive to the telluric correction. While the
blueward rise begins well before this difficulty, the exact form of
the slope of the continuum below 2.0 $\mu$m is not well-defined with the
current data set.

\subsection{The Secondary Star of EF Eri}

As we have just noted (see Fig. 5), combining the entire NIRSPEC data
set has allowed us to tentatively detect the secondary star in EF Eri. While
the S/N of these data are not high, if the secondary star were relatively
normal, it should be more prominent than seen here. 
By modeling the optical spectrum of EF Eri, Beuermann et al. 
(2000) have derived distances for EF Eri in the range 80 $\leq$ d $\leq$
128 pc, while Thorstensen (2003) has measured a distance of 163 $\pm$ 58 pc
via parallax. Thorstensen finds that an astrometric solution with a distance 
of 113 pc is possible by adjusting some of the input criteria. 
As shown in Harrison et al. (2003), the optical photometry of EF Eri is 
well-modeled by a 9500 K white dwarf, and this white dwarf provides no more 
than 10\% of the flux in the {\it K}-band. If we assume that 
the secondary star is responsible for the remainder of the {\it K}-band flux 
at phase 0.0 ({\it K} = 15.55), the absolute magnitude of EF 
Eri is in the range 10.0 $\leq$ M$_{\rm K} \leq$ 11.0. This absolute 
magnitude corresponds to a spectral type in the range M8 to L1 (ignoring any 
effects due to the Roche lobe filling nature of this object).

A secondary star with a spectral type near M9 should have strong
CO features, as well as a detectable Na I doublet. We have compared the 
combined {\it H}- and {\it K}-band spectra of EF Eri with published spectra
for stars and brown dwarfs in the range M7V to T8. While the overall 
slope of the continuum of the phase 0.0 {\it K}-band spectrum
is consistent with a very late-type M dwarf (see Fig. 6), the {\it H}-band
spectrum bears no resemblence to such an object. The departure 
becomes more pronounced when we compare the spectrum of EF Eri 
to those of L-dwarfs. While brown dwarfs around L5 show a similar spectral 
slope at
1.55$-$1.6\,$\mu$m, there is no feature comparable to the strong flux
peak at 1.7\,$\mu$m. While the absence of Na\,I in the $K$-band is 
consistent with a spectral type later than L0 (McLean et al. 2003), the
water vapor absorption bands of such an object are much deeper and
depress the blue side of the $K$-band spectrum much further to the red
($>$\,2.1\,$\mu$m) than is seen in EF Eri. Also, the first overtone CO bandhead at 2.29\,$\mu$m 
remains strong in brown dwarfs until CH$_4$ takes over in the late L types.

In addition to the difference in $H$- and $K$-band spectral features,
the $JHK$ photometry is almost impossible to reconcile with any known
dwarf. While the {\it H} to {\it K}-band flux ratios for
EF Eri are more-or-less consistent with those of late type objects, giving
rise to ($H - K$) = 0.42 ($\approx$ M8V), the observed ($J  -  K$) color 
($>$ 2.5) exceeds even the reddest IR colors observed in mid-L
dwarfs (Legget et al. 2002). 

\subsubsection{Model Spectra for Carbon/Oxygen-depleted Atmospheres}

As shown in Harrison et al. (2004), many CV secondary stars show evidence
for weak CO absorption features. Since the water vapor features in those
objects appear to be relatively normal, Harrison et al. suspect that
Carbon is deficient in those systems.  This view is supported by analysis
of FUV observations of CVs by G$\ddot{a}$nsicke et al. (2003) where 
large N V/C IV ratios have been found, suggesting Nitrogen enhancements, and 
Carbon
deficits in the photospheres of the white dwarf primaries, presumably
arising from matter transferred from the secondary star. It would therefore
not be unexpected for the secondary star in EF Eri to have anomalously
weak CO absorption features. A lowered Carbon abundance would also reduce the
CH$_4$ absorption expected for late-L and T dwarfs, and thus could 
produce much redder ($J\!-\!K$) colors through the range of
possible spectral types for the secondary of EF Eri.

To test this hypothesis we have calculated theoretical spectra using
the general stellar atmosphere code \texttt{PHOENIX} (Hauschildt et al. 1999)
for a variety of chemical compositions that might be expected in a
CNO-processed stellar core.
Depth-dependent partial pressures for more than 650 species are
calculated using the EOS of Allard \& Hauschildt (1995) by solving chemical
equilibrium for 40 elements, as described in Allard et al. (2001), including 
dust formation.
Model structures and flux spectra are computed self-consistently,
treating line blanketing by the direct opacity sampling method, with
line strengths taken from a master database of 43 million atomic and 700 
million molecular lines. In addition, dust extinction profiles for 31 grain
species are included.
Major updates relative to the AMES-Cond and AMES-Dusty models
of Allard et al. (2001) are the addition of 30 million methane lines from
Homeier et al. (2004a), and the explicit treatment of the gravitational settling
of dust grains developed by Allard et al. (2003).
In addition, the Ames list of water line opacities from
Partridge \& Schwenke (1997) has been replaced by the earlier list from
Miller et al. (1994) for these models. The latter
list is less complete for high temperature steam, but has been found
to give better agreement with the IR spectra of cool brown dwarfs
(Homeier et al. 2004b).
Nevertheless we emphasize that none of the currently available
databases for molecular opacities accurately reproduces the observed
shape of all absorption bands in the IR. These models also suffer from
incomplete or missing data for several molecules, such as CrH
and FeH. For example, the $H$-band spectra of mid-L dwarfs show
a moderately strong absorption feature near 1.6\,$\mu$m (see Fig. 6) that
is not reproduced by any of our models. Several features in this
region have been identified by Cushing et al. (2003) with the
$E^4\Pi$--$A^4\Pi$ system of FeH, for which no line opacity data are
available at this time.
Possibly due to this missing opacity source, and other uncertainties
in input data such as H$_2$O opacities and dust extinction
profiles, our theoretical spectra tend to produce
($H\!-\!K$) colors that are too blue when compared to observed L dwarfs
(Legget et al. 2001).

We have constructed grids of models based on revised solar abundances
as compiled by Lodders (2003), except for Carbon, Oxygen and
Nitrogen, which were altered as indicated for the individual
models. The basic atmospheric parameters span the range
1400~K $\leq$ T$_{\rm eff}$ $\leq$ 2500~K, and
log\,$g$\,=\, (4.5, 5.0, 5.5). First, investigating the effects of
lowering the Carbon abundance to [C/Fe]\,=\,($-$0.5, $-$1.0, $-$1.5),
we find that a depletion to at least 0.1\,$\times$\,solar is required
to explain the weakness or absence of the CO overtone bands.
 From synthetic photometry we can also exclude models with
$T_{\rm eff}$ significantly higher than 2000\,K, as these disagree
with the observed $J$ magnitudes even assuming the upper limit of 5.5
for the distance modulus. Below 1600\,K, the strong methane
bands at 2.2\,--\,2.3\,$\mu$m become prominent even at
[C/Fe]\,=\,$-$1.0, but lowering the Carbon abundance further does not
improve the fit to the observed $K$-band spectrum  at
such low temperatures. We find best agreement for the red part of the
$K$-band spectrum for models with 1600 $\leq$ T$_{\rm eff}$ $\leq$ 1800\,K, as 
shown by the
bottom spectrum in Fig. 7. But a Carbon deficiency alone cannot explain the
early onset of absorption to the blue side of the $K$-band, which appears at
much shorter wavelengths than typically seen in both our models and
in observations of L dwarfs. Nor can a lack of Carbon be responsible for
the peculiar shape of the $H$-band spectra. The
main opacity source at these wavelengths is water vapor, and
significant changes can only be expected if the Oxygen abundance is
reduced as well. Such effects might be expected if the secondary CNO
cycle has been active for some time, though models by Marks \&
Sarna (1998) do not indicate significant changes in the Oxygen
abundance for CNO processed material. We have nonetheless produced test models 
for [O/Fe]\,=\,$-$0.5 to explore this case.
As one would expect, these models (Fig. 7) show
an onset of the water vapor bands both in the $H$ and $K$-band at
shorter wavelengths, with a $K$-band spectrum that fits our
observations quite well. For comparison, a model with additionally
enhanced Nitrogen abundance, as would be expected with the depleted
Carbon and Oxygen transformed into $^{14}$N, is also shown. 

Synthetic
colors calculated for these models show that the Nitrogen-enriched
atmosphere produces the closest match to the ($J\!-\!K$) and
($H\!-\!K$) photometry. We have not attempted
to modify our input chemistry further to optimize these fits, since
such results would necessarily be of limited significance given the
uncertainties in the available opacity data and the limited wavelength
coverage of our spectra. Clearly a direct detection of the water absorption in
the $H$-band, which we expect to occur at 1.5\,--\,1.55\,$\mu$m, and
$J$-band spectra, would help in resolving these uncertainties. Such data
would also allow for the study of irradiation, which is expected to have
the strongest impact at shorter wavelengths (Barman et al. 2000).
In conclusion, a brown dwarf with a considerable deficiency of Carbon,
a lesser depletion of Oxygen, and an enhancement of Nitrogen,
reproduces much of the secondary's IR SED, and the shape of the
$K$-band at phase 0.0. The additional absorption at 1.6\,$\mu$m cannot be 
accounted for in the current models. If this band is indeed due to FeH, we 
expect it to become relatively stronger when the steam bands are reduced.
In that case the spectra could probably reproduce the ``base'' of
the $H$-band emission. The shape of the actual peak, however, does not
correspond to any known stellar features, and also shows changes with
phase that cannot be explained by irradiation effects.
We conclude that cyclotron emission at 1.6\,--\,1.8\,$\mu$m
contaminates the infrared spectrum for all orbital phases.

\subsection{The Cyclotron Features}

It is clear that the {\it H}-band spectra are inconsistent with that of a 
normal (or chemically altered) late-type or brown dwarf secondary star. 
If the phase 0.0 dip near 1.6 $\mu$m is associated with FeH, then we estimate
that $\approx$ 30\% of the $H$-band flux at this juncture is due to cyclotron
emission.  Thus, to produce the observed, phase-dependent changes in
the {\it H}-band spectra probably requires two separate cyclotron features 
in this bandpass. If we assume the phasing and geometry from 
Bailey et al. (1982), the phase 0.0 cyclotron features that we see in the
{\it H}-band are 
presumably associated with the pole that accretes during the high state. This 
suggests that the single, dominant cylcotron feature at phase 0.5 might be from 
the other pole.

While the minimum light {\it K}-band spectrum is fairly consistent with 
the shape of a cool star spectrum, with little evidence for cyclotron
emission, the maximum light spectrum is clearly contaminated
by at least one, possibly two cyclotron features. Obviously, there is a 
cyclotron harmonic that peaks below 2.0 $\mu$m. The rapid change in slope of
the red half of the $K$-band spectrum, and the dip near 2.07 $\mu$m, suggest
the possibility of an additional, weaker cyclotron feature at phase 0.5. 
To better examine the phase 0.5 cyclotron features, we have subtracted the 
phase 0.0 spectrum from the phase 0.50 spectrum, to produce the residual 
spectrum shown in Fig. 8. The result shows two strong cyclotron features, one
that peaks in the {\it H}-band near 1.58 $\mu$m, and one that (we estimate) 
peaks near 1.93 $\mu$m. The possibility of a second cyclotron feature in the
{\it K}-band seems more likely after this subtraction. If the secondary star
has a similar spectrum at phases 0.0 and 0.5, then any water vapor feature
should disappear upon subtraction. Ignoring the false H I Br-$\gamma$ emission
feature, there continues to be a modest dip near 2.07 $\mu$m and a small
rise that peaks near 2.15 $\mu$m and that falls off slowly beyond 2.2 $\mu$m. 

\section{Discussion}
Regardless of the exact interpretation of the spectra presented here, it is
clear that the model proposed by Harrison et al. (2003) is incorrect. The
photometric modulations in the {\it K}-band are certainly being driven by the 
changing strength of the cyclotron feature at the blue end of this bandpass
(and possibly aided by a broader, and weaker feature that peaks near 2.15 
$\mu$m).  There is very little evidence for the secondary star in our 
{\it K}-band spectra, with only the suggestion of a weak first overtone CO 
feature in the median of the {\it K}-band data set.
While the presence of weak CO features in a CV is not surprising (see Harrison 
et al.  2004), stars with the late spectral type expected for the secondary in 
EF Eri have {\it H}-band spectra that are completely different to what
we have observed. While model brown dwarf spectra with dramatic alterations in 
the abundances of the major opacity sources at these wavelengths do a somewhat
better job at fitting the data, neither the observations or the models 
are currently adequate enough to truly investigate the nature of the underlying
secondary star. 

The fact that cyclotron emission can still dominate the spectrum of EF Eri
in this low state is surprising given that EF Eri exhibits no outward signs of 
continued accretion. The optical and infrared spectra lack Balmer 
and Brackett emission from H I. The optical {\it BVRI} spectral energy 
distribution and spectrum are completely consistent with that of an inactive
magnetic white dwarf. The model white dwarf atmosphere, with T$_{\rm eff}$ =
9500 K, used by Harrison et al. (2003) to compare with their 2001 March 11
spectrum of EF Eri, does an excellent job at fitting the 2002 August 10 optical
spectrum we obtained with the NTT. Clearly, the white dwarf photosphere alone 
is too cool to supply significant quantities of ionizing photons. It appears
that the white dwarfs in ultra-short period, low mass transfer rate CVs are
significantly cooler than those in longer period, or higher mass transfer rate
systems (c.f., Szkody et al. 2002).

Four magnetic CV systems have now been found that display a single,
dominant cyclotron feature in their optical spectra that have
been attributed to very low mass transfer rates: SDSS 
J155331.12+551614.5 and 
SDSS J132411.57+032050.5 (Szkody et al. 2003), RBS0206 (= CQ Cet; Schwope, 
Schwarz, \& Greiner 1999), and HS 1023+3900 (= WX LMi; Reimers, Hagen \& 
Hopp 1999). Along with the single very large cyclotron feature is evidence for
a second, weaker feature at the next lower/higher harmonic number covered by
their optical spectra. Of these four polars, both SDSS J155331.12+551614.5 and 
J132411.57+032050.5 have optical light curves that resemble the
infrared light curves of EF Eri (see Figs. 2 and 6 of Szkody et al. 
2003). SDSS1553 shows optical variations with peak-to-peak amplitudes of
$\Delta$m $\approx$ 1 mag in {\it V} and {\it R}, while SDSS1324 has variations 
in the Sloan {\it r}-band of $\Delta r$ = 1.3 mags. Both sets of light curves
are roughly sinusoidal, like the {\it H} and {\it K}-band light curves of
EF Eri. Unlike EF Eri, however, the spectra of both SDSS1553 (Fig. 1 in
Szkody et al. 2003) and SDSS1324 (G. Schmidt, priv. comm.) reveal H I emission.

The interpretation of the spectra of SDSS1553, SDSS1324, CQ Cet and WX LMi
is that they have entered phases of extremely low accretion rates,
the so-called ``bombardment solution'' (Kuijpers \& Pringle 1982), where any 
heating is from particle collisions instead of shocks. This state
is characterized by low plasma temperatures, and mass transfer rates
of \.{M} $\leq$ 10$^{\rm -13}$ M$_{\sun}$ yr$^{-1}$. In each of these
four polars, the harmonic numbers of the dominant cyclotron features are in the
range 2 $\leq  n  \leq$ 4, with magnetic field strengths above 45 MG.
Magnetic field strength estimates for EF Eri
range from 10 to 21 MG. Assuming B $\sim$ 14 MG, 
the $n$ = 4 harmonic would peak at 1.96 $\mu$m, and the $n$ = 5 harmonic 
would peak at 1.58 $\mu$m. The $n$ =3 harmonic would peak near 2.6 $\mu$m,
while the $n$ = 6 and 7 harmonics would appear in the {\it J}-band. It
is worth noting that during the {\it H}- and {\it K}-band maxima, the 
{\it J}-band light curve shows a minimum. Given this fact, and the need
for very high accretion rates to produce power in these higher harmonics,
it is likely that the $n$ = 6 and 7 harmonics are not significant sources of 
flux in EF Eri. The fact that the {\it J}-band light curve has a maximum 
during the {\it H}- and {\it K}-band 
minima suggests activity from another pole with a significantly higher
magnetic field strength. The flux peak at 1.72 $\mu$m, and the upturn at the 
blue end of the phase 0.0 {\it H}-band 
spectra, along with the lack of significant cyclotron emission in the $K$-band 
indicate that the phase 0.0 cyclotron emission comes from a region with a 
significantly different field strength.  A fit to the cyclotron features of
the high state $JHK$ spectra of EF Eri by Ferrario et al. (1996) suggested 
emission from two regions with magnetic field strengths of 16.5 and 21 MG.
An accreting pole with a field strength of 21 MG, would not produce a 
cyclotron feature in the $K$-band, but would produce harmonics at 1.7 and 
1.3 $\mu$m, consistent with our phase 0.0 spectra.

That the magnetic field structure of EF Eri is complex, including strong
evidence for emission from more than one pole, has been demonstrated by 
Hutchings et al. (1982), Piirola et al. (1987), Meggit \& Wickramasinghe
(1989), and Ferrario et al. (1996).
Recent ``magnetic field topology'' mapping of the magnetic field of the
white dwarf in EF Eri by Reinsch et al. (2003) reveals this complexity:
there are two main polar caps with field strengths as high 100 MG
at their centers, as well as evidence for multiple, zonal components.
The complex structure of the field in EF Eri must be partly to blame
for the results presented here. Modeling the cyclotron features that would
arise from accretion onto such a field would be helpful in unraveling our
spectroscopic results.

The lack of detectable H I emission from EF Eri, when compared to the other
members of its small family, is interesting. It is highly likely 
that the current accretion rate in EF Eri is lower than the other four 
objects, and is simply not sufficient to supply the heating necessary to 
generate the UV photons required to create H I emission. In the bombardment
scenario, the accretion shock disappears, and the white dwarf photosphere
is directly heated by particle collisions. A lower accretion rate means less 
heating, and therefore less ionizing radiation.

The secondary star of EF Eri remains elusive. Beuermann et al. (2000)
have presented arguments that the spectral type of the secondary star
in EF Eri must be later than M9, with the most probable spectral type
later than L3. We can estimate the maximum amount of flux from the
secondary by assuming that the flux level of the dip seen in the phase 0.0 
{\it H}-band spectrum (at 1.57 $\mu$m) represents the continuum level of
the underlying secondary star. This corresponds to {\it H}$_{\rm sec} \geq$ 
16.5. Assuming 80 $\leq$ d $\leq$ 128 pc, the absolute {\it H} magnitude 
of the secondary star is in the range 11.0 $\leq$ M$_{\rm H}$ $\leq$ 12.0,
consistent with an early/mid-L dwarf. Thus, EF Eri joins WZ Sge (see Howell
et al. 2004) as the cataclysmic variables with the strongest evidence for 
harboring brown dwarf-like secondary stars.

\section{Conclusions}

We have obtained phase-resolved infrared spectroscopy of EF Eri and conclude
that the large amplitude variations seen in its infrared light curve are not 
due to a heated brown dwarf, but has its origin in cyclotron emission. EF Eri
appears to join the small family of polars where the accretion rate has 
dropped to a very low level. 
 Unfortunately, our data are inadequate to allow us unravel the 
complex cyclotron emission that is present in EF Eri. 
Phase-resolved {\it J}-band spectroscopy is required to make further progress
on disentangling the cyclotron spectrum of EF Eri. 
Given that the mean {\it J}-band magnitude of EF Eri is 17.3, multiple
orbits of data on 8 m-class telescopes will be needed to obtain spectra with
sufficient S/N. To fully interpret the spectroscopic observations,
phase-resolved, broad-band {\it JHK} polarimetric observations of EF Eri will 
be essential.  
It would also be extremely useful to have a radial velocity curve for the EF 
Eri system to allow us to properly phase, and coadd the infrared spectra to 
search for features from the secondary star. Such
data could be acquired through optical spectroscopy of the 
Zeeman absorption features of H I. 

\acknowledgements{TEH acknowledges partial support under NSF grant
AST 99-86823. We would like to thank T. Geballe for his help in planning
and executing our NIRI observations, and R. Campbell and G. Wirth for
assistance with NIRSPEC. DH acknowledges support from the National
Science Foundation under NFS grant N-Stars RR185-258, and from NASA
under grant NLTE RR185-236. Models presented in this work are based in part on 
calculations performed at the NERSC IBM SP at LBNL with support from the DoE.
We would also like to thank F.~Allard and P.~Hauschildt for assistance in
construction of the model atmospheres, T.~Barman for helpful discussions
on the effects of irradiation, and I.~Baraffe for providing us with results of
evolutionary calculations for interacting brown dwarfs. We also would like
to thank our referee D. Wickramasinghe, and editor J. Leibert for additional
insight on EF Eri, and for suggesting changes to improve the clarity of our
paper.  Based partly on observations obtained at the Gemini 
Observatory, which is operated by the Association of Universities for Research 
in Astronomy, Inc., under a cooperative agreement with the NSF on behalf of 
the Gemini partnership: the National Science Foundation (United States), the 
Particle Physics and Astronomy Research Council (United Kingdom), the
National Research Council (Canada), CONICYT (Chile), the Australian Research 
Council (Australia), CNPq (Brazil) and CONICET (Argentina). Additional
data presented herein were obtained at the W.M. Keck Observatory, which is 
operated as a scientific partnership among the California Institute of 
Technology, the University of California and the National Aeronautics and 
Space Administration. The Observatory was made possible by the generous 
financial support of the W.M. Keck Foundation. The authors wish to recognize 
and acknowledge the very significant cultural role and reverence that the 
summit of Mauna Kea has always had within the indigenous Hawaiian community.  
We are most fortunate to have the opportunity to conduct observations from 
this mountain.

\begin{flushleft}
{\bf References}

Allard, F. \& Hauschildt, P.~H. 1995, ApJ, 445, 433\\
Allard, F., Hauschildt, P., Alexander, D., Tamanai, A., \& Schweitzer, A. 
  2001, \apj, 556, 357\\
Allard, F., Guillot, T., Ludwig, H., Hauschildt, P.~H., Schweitzer,
A., Alexander, D.~R., \& Ferguson, J.~W. 2003, in IAU Symposia, Vol. 211,
{\it Brown Dwarfs}, ed. E.~Mart{\'i}n, International Astronomical Union (San
Francisco: Astronomical Society of the Pacific), p. 325\\
Bailey, J., Hough, J. H., Axon, D. J., Gatley, I., Lee, T. J., Szkody, P.,
Stokes, G., \& Berriman, G. 1982, MNRAS, 199, 801\\
Barman, T. S., Hauschildt, P.~H., \&  Allard, F. 2002, in ASP Conf. Ser.
 261: The Physics of Cataclysmic Variables and Related Objects, (San
 Francisco: Astronomical Society of the Pacific), p49\\
Beuermann, K., Wheatley, P., Ramsay, G., Euchner, F., \& Gansicke, B. T.
2000, A\&A, 354, L49\\
Cushing, M.~C., Rayner, J.~T., Davis, S.~P., \& Vacca, W.~D. 2003,
\apj, 582, 1066\\
Ferrario, L., Bailey, J., \& Wickramasinghe, D. 1996, MNRAS, 282, 218\\
Harrison, T. E., Howell, S. B., Huber, M. E., Osborne, H. L., Holtzman, J. A.,
Cash, J. L., \& Gelino, D. M. 2003, AJ, 125, 2609\\
Harrison, T. E., Osborne, H. L., Howell, S. B., 2004, AJ, 127, 3493\\
Hauschildt, P. H., Allard, F., \& Baron, E. 1999, ApJ, 512, 377\\
Homeier, D., Allard, F., Hauschildt, P., \& Boudon, V. 2004a, \apj, in prep.\\
Homeier, D., Burgasser, A., Hauschildt, P., Allard, F., Allard, N.,  
  McLean, I., McGovern, M., Kirkpatrick, J., \& Prato, L. 2004b, \apj, to be  
  submitted \\
Howell, S. B., Harrison, T. E., \& Szkody, P. 2004, ApJ, 602, L49\\
Howell, S. B., Nelson, L. A., \& Rappaport, S. 2001, ApJ, 550, 897\\
Hutchings, J. B., Cowley, A. P., Crampton, D., Fisher, W. A., \& Liller, M. H. 
1982, ApJ, 252, 690\\
Kuijpers, J., \& Pringle, J. E. 1982, A\&A, 114, L4\\
Lancon, A., \& Rocca-Volmerange, B. 1992, A\&A Supp., 96, 593\\
Leggett, S.~K., Allard, F., Geballe, T.~R., Hauschildt, P.~H., \&
  Schweitzer, A. 2001, \apj, 548, 908\\
Lodders, K. 2003, ApJ, 591, 1220\\
Maiolino, R., Rieke, G. H., \& Reike, M . J. 1996, AJ, 111, 537\\
Marks, P. B., \& Sarna, M. J. 1998, MNRAS, 301, 699\\
McLean, I.S., McGovern, M.R., Burgasser, A.J., Kirkpatrick, J.D., Prato, L., 
  \& Kim, S.S. 2003, ApJ, 596, 561\\
Meggitt, S. M. A., \& Wickramasinghe, D. T. 1989, MNRAS, 236, 31\\
Miller, S., Tennyson, J., Jones, H., \& Longmore, A. 1994, in IAU  
  Colloquia, Vol. 146, Molecular Opacities in the Stellar Environment, ed.  
  P.~Thejll \& U.~G. J{\o}rgensen, International Astronomical Union (Niels Bohr 
  Institute and Nordita press, Copenhagen), 296\\
Partridge, H. \& Schwenke, D.~W. 1997, \jcp, 106, 4618\\
Piirola, V., Reiz, A., \& Coyne, G. V., 1987, A\&A, 186, 120\\
Reimers, D., Hagen, H., \& Hopp, U. 1999, A\&A, 343, 157\\
Reinsch, K., Euchner, F., \& Beuermann, K. 2003, astro-ph/0302056\\
Schwope, A. D., Schwarz, R., \& Greiner, J. 1999, A\&A, 348, 861\\
Szkody, P. et al. 2003, ApJ, 583, 902\\
Szkody, P., Gansicke, B. T., Sion, E. M., \& Howell, S. B. 2002, ApJ, 574, 950\\
Thorstensen, J. R. 2003, AJ, 126, 3017\\
Wickramasinghe, D. T., \& Ferrario, L. 2000, PASP, 112, 873\\
\end{flushleft}

\begin{figure}
\epsscale{0.95}
\plotone{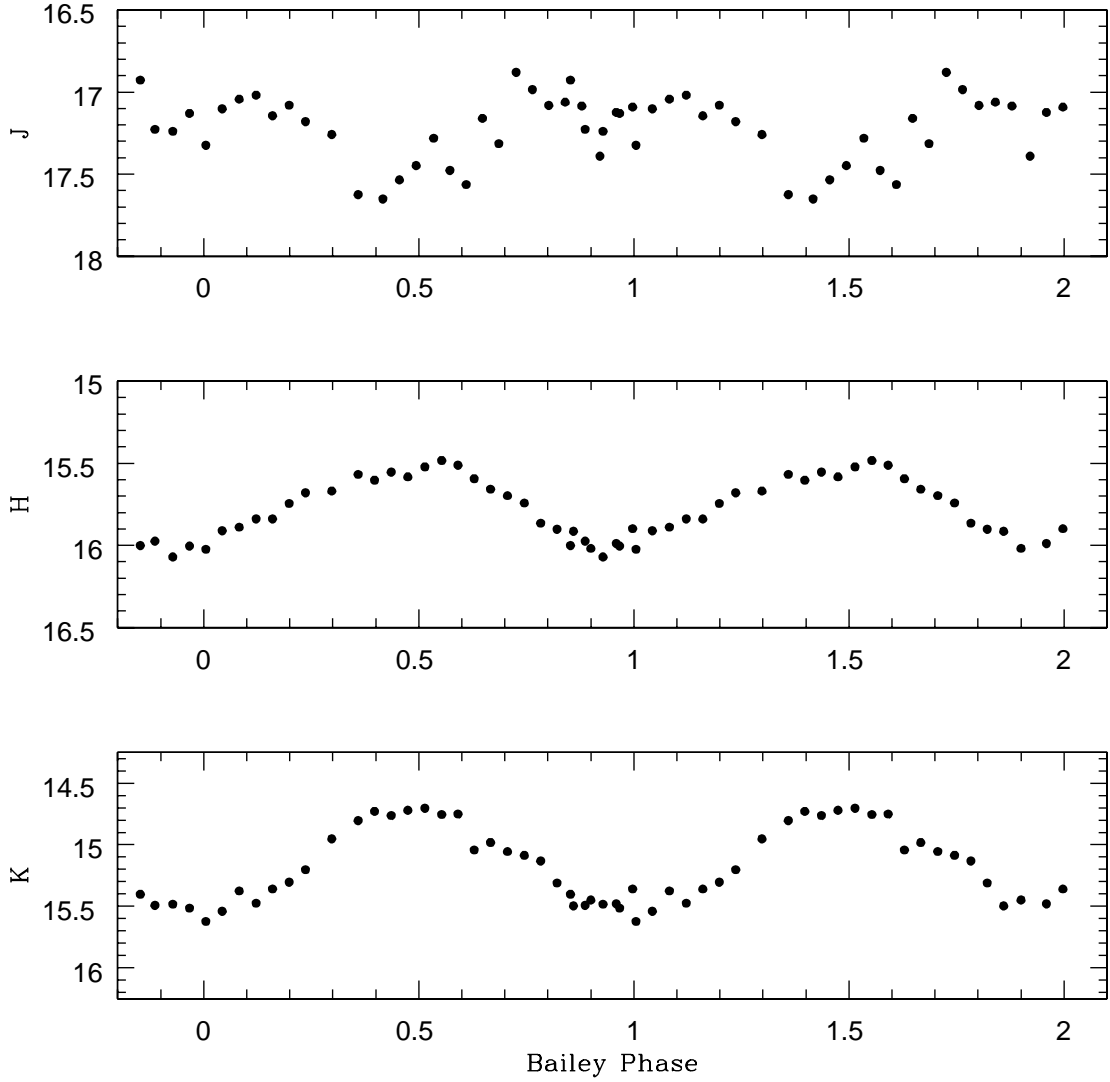}
\caption{The infrared light curves of EF Eri obtained with SQIID on the
KPNO 2.1 m. The absolute fluxes and light curve amplitudes have not changed 
when compared to those presented in Harrison et al. (2003). The data
sets have been plotted over two orbital cycles for clarity.}
\end{figure}

\begin{figure}
\epsscale{0.85}
\plotone{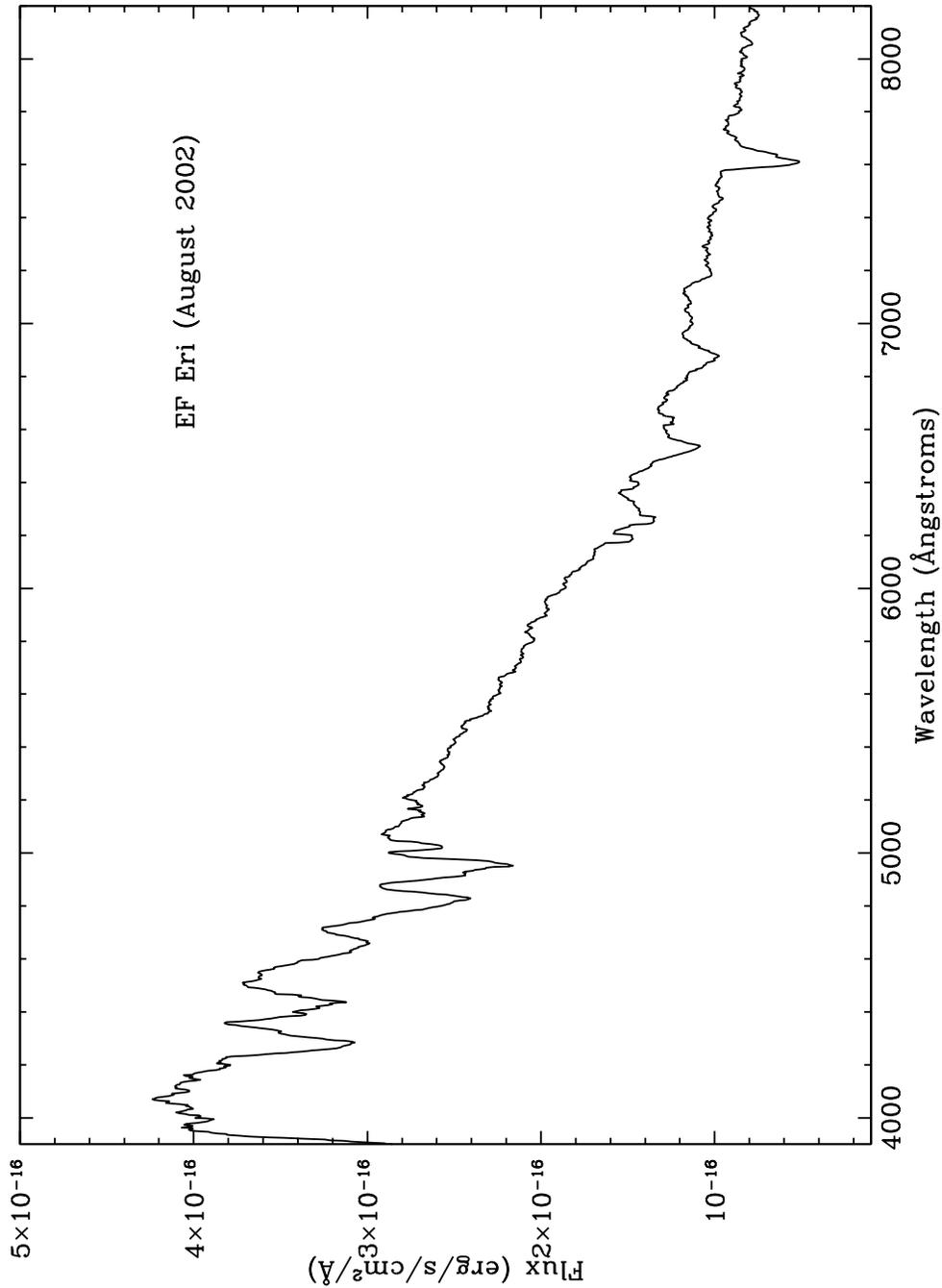}
\caption{An optical spectrum of EF Eri obtained with EMMI on the {\it NTT} on 
2002 August 10. The
accretion rate in EF Eri has declined to the point where there is no longer
detectable emission at H$\alpha$. The $\sigma^{\pm}$ features from the Zeeman
splitting of the H I lines, however, are clearly seen. The central,
unshifted absorption components of H I are weaker than expected, suggesting
the possibility of continued low-level emission.}
\end{figure}

\begin{figure}
\epsscale{0.90}
\plotone{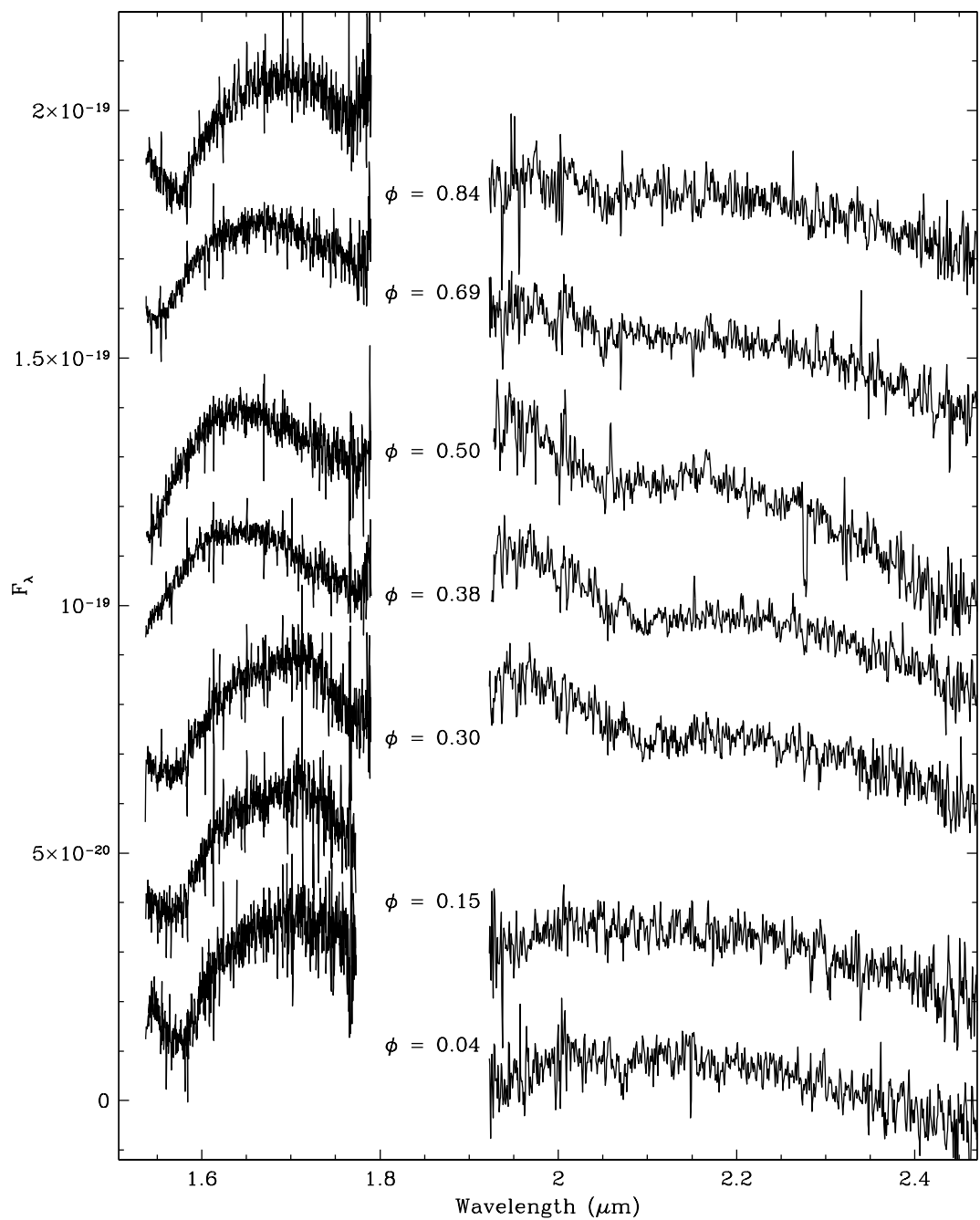}
\caption{The Keck NIRSPEC {\it H-}band, and Gemini NIRI {\it K}-band, spectra
of EF Eri. Both data sets have been flux-calibrated using the infrared
light curves. For clarity, the {\it H}- and {\it K}-band spectra for each 
orbital phase have been offset in flux. The mean orbital
phases of the spectra are listed.}
\end{figure}

\begin{figure}
\epsscale{0.95}
\plotone{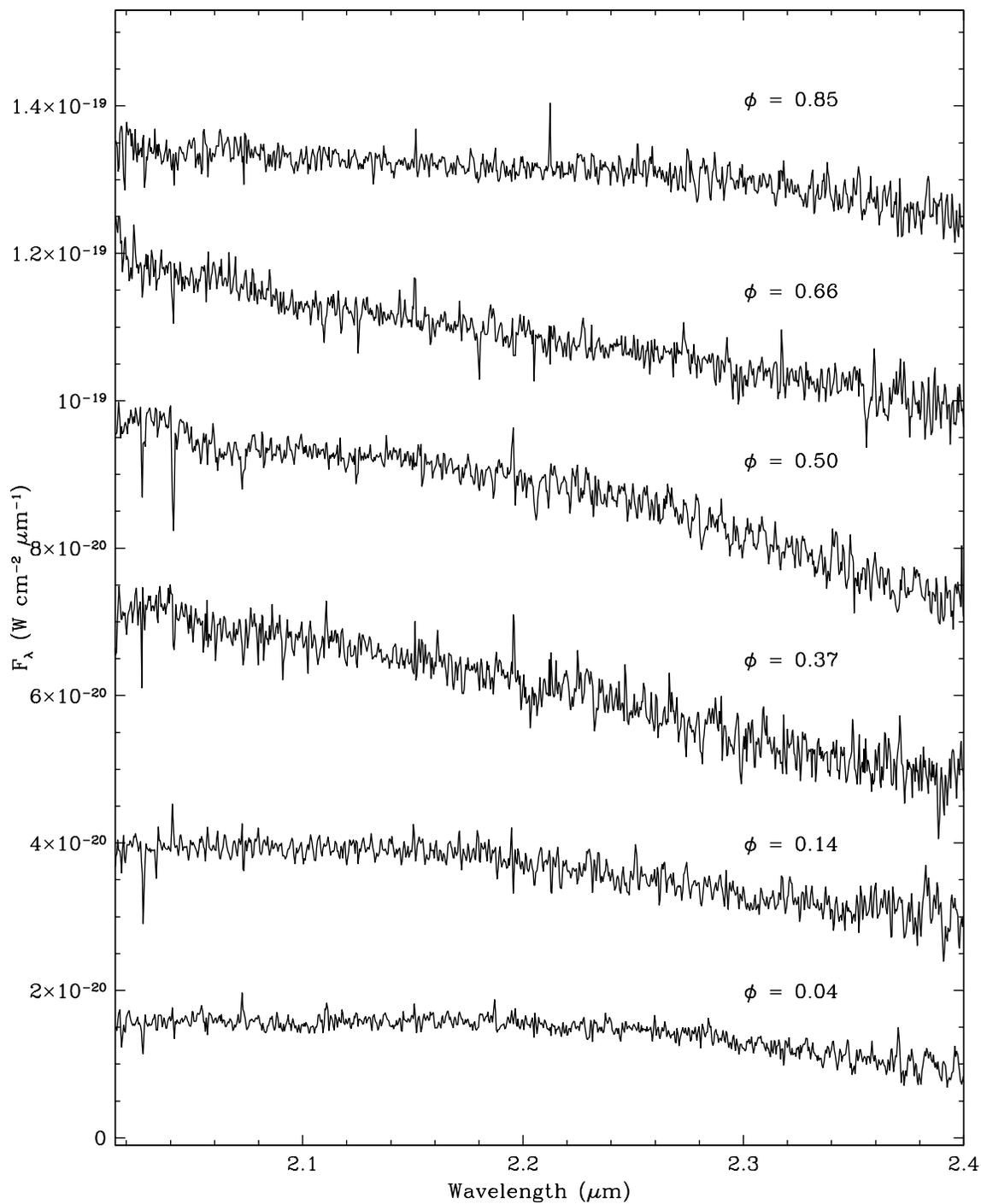}
\caption{The NIRSPEC {\it K}-band phase-resolved spectra of EF Eri. These
data have been flux-calibrated using the infrared light curves and then
offset for clarity. The same phase-dependent morphology seen in the Gemini 
data set is present here. Each spectrum is the median of several
three minute exposures.}
\end{figure}

\begin{figure}
\epsscale{0.95}
\plotone{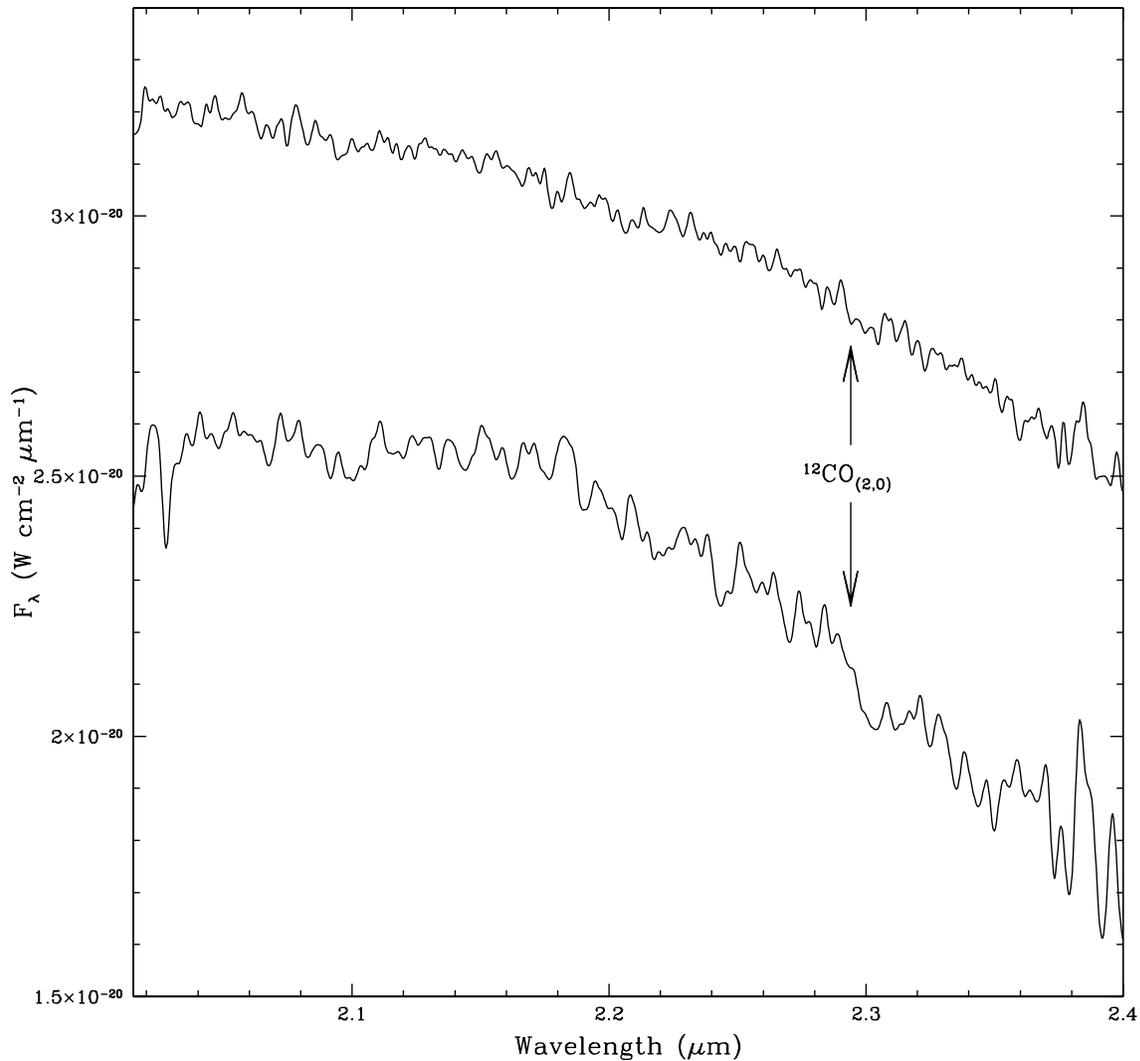}
\caption{The median of the entire NIRSPEC {\it K}-band data set (top, offset
in flux),
and a smoothed (to FWHM = 60 \AA) median of the seven spectra closest to phase 
0.0 (bottom). A possible, very weak
absorption feature at the position of the first overtone of 
$^{\rm 12}$CO$_{\rm (2,0)}$  (at 2.294 $\mu$m) is seen in both spectra,
and appears to be the only indication of the presence of the secondary star, 
besides the decline at the red end of the {\it K}-band that might be associated with water vapor 
absorption.}
\end{figure}

\begin{figure}
\epsscale{0.95}
\plotone{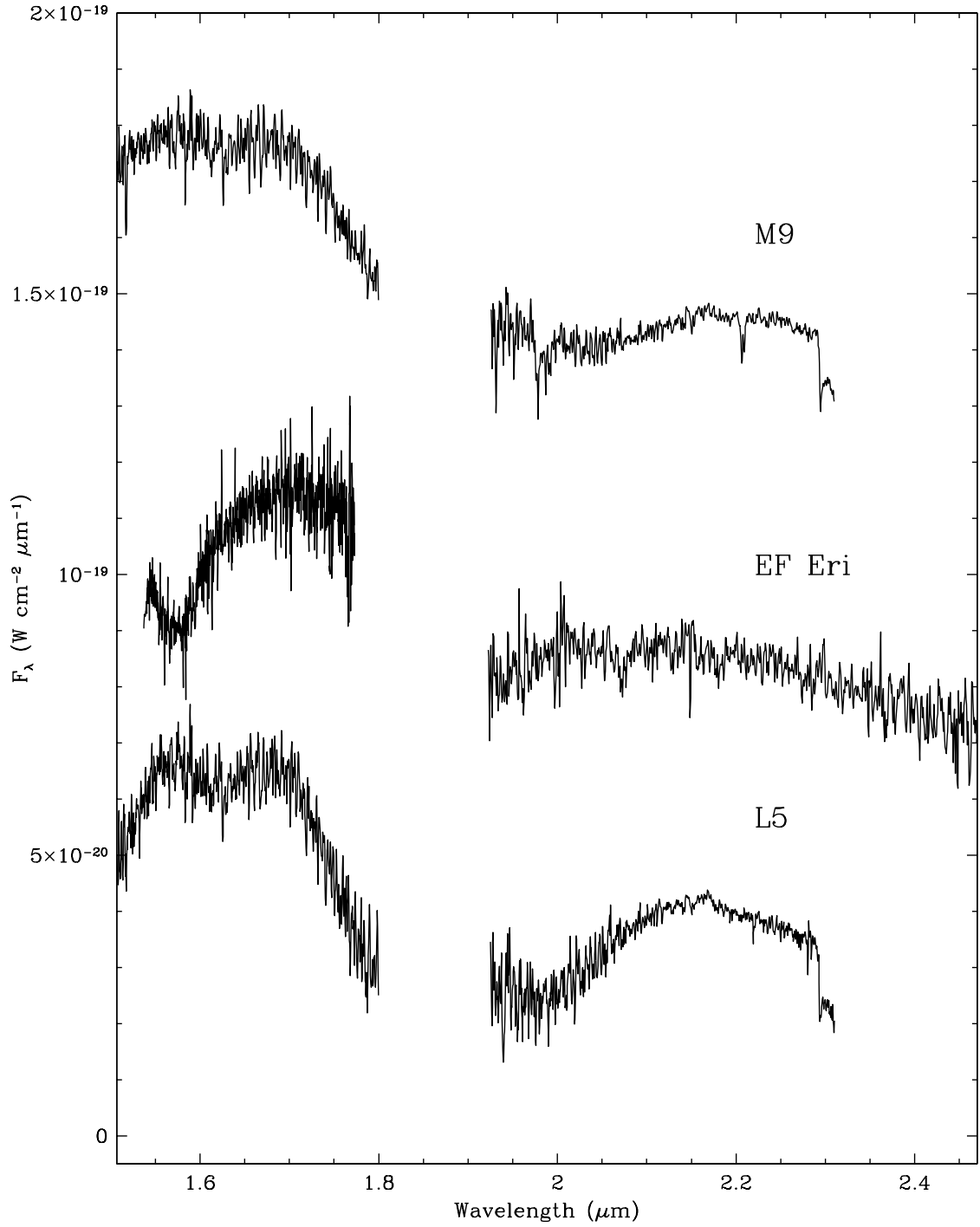}
\caption{Comparison of the phase 0.0 spectrum of EF Eri to those of
M9 and L5 dwarfs (from McLean et al. 2003). The sharp feature at 2.294 $\mu$m
in the comparison spectra is due to the first overtone feature of CO.}
\end{figure}

\begin{figure}
\epsscale{0.80}
\plotone{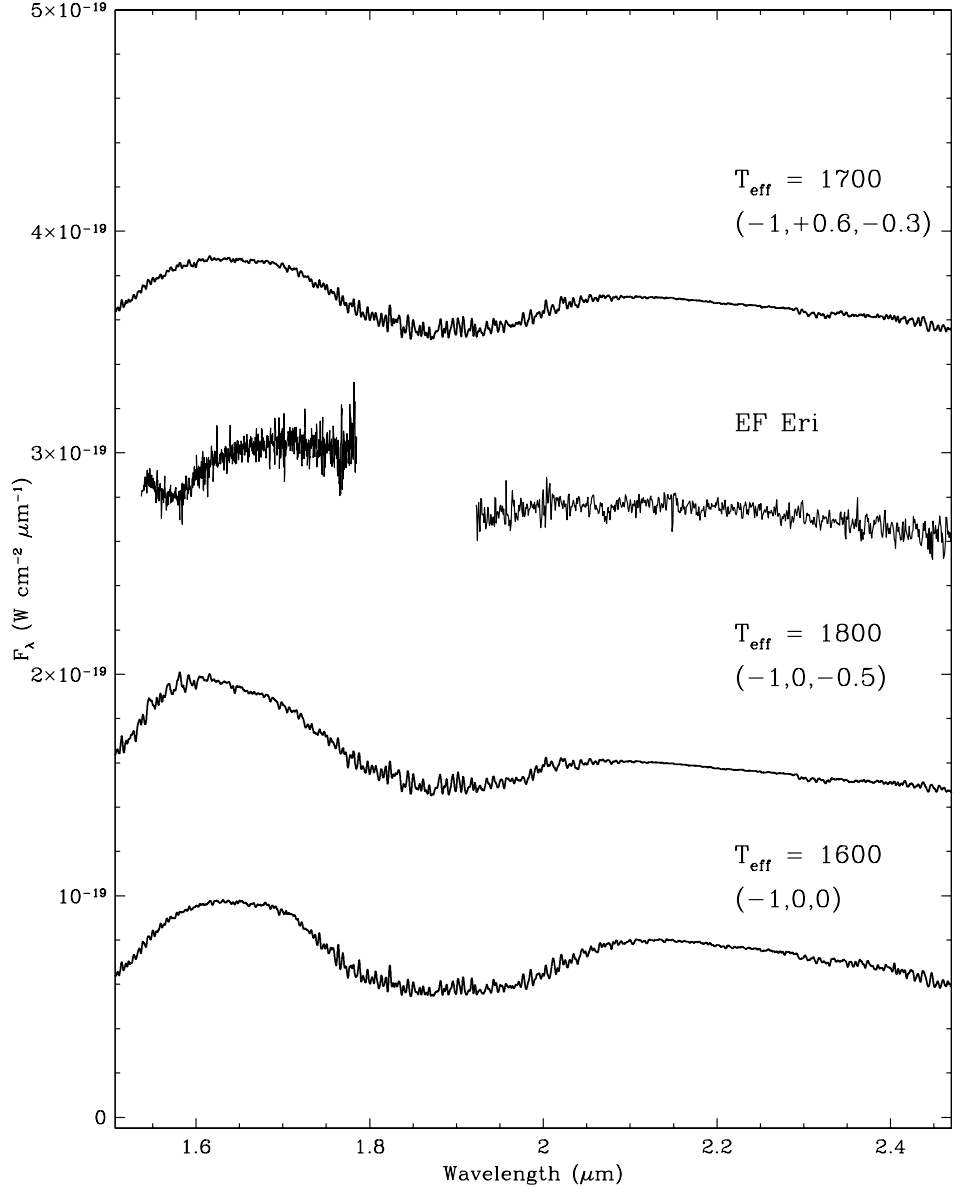}
\caption{Comparison of the phase 0.0 spectrum of EF Eri to brown dwarf model
spectra generated using \texttt{PHOENIX}. The temperature of each model is listed,
and each of these models has a gravity of log{\it g} = 5.0.
Below the temperature are the abundances of Carbon, Nitrogen, and Oxygen (in
dex, vs. solar). The top model has deficits of Carbon and Oxygen, and enhanced
levels of Nitrogen (4$\times$ solar). The bottom-most model simply has
a Carbon deficit (0.1$\times$ solar). The Oxygen deficient models fit the
observed {\it K}-band spectrum due to their shallower water vapor absorption.
None of the models we developed, however, could explain the shape of the
{\it H}-band spectrum of EF Eri.}
\end{figure}

\begin{figure}
\epsscale{0.95}
\plotone{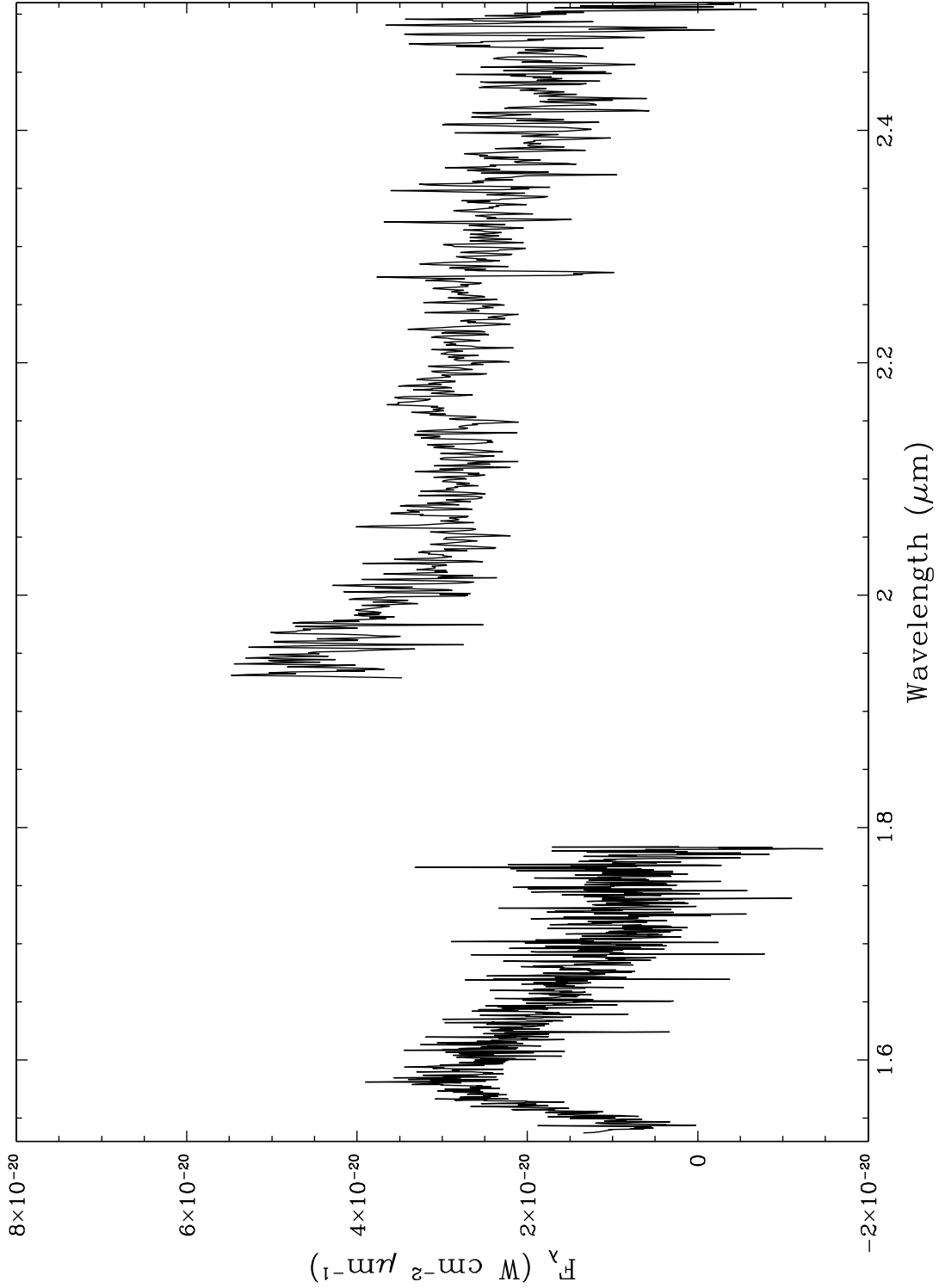}
\caption{The residual spectrum created by subtracting the phase 0.0 spectra
from the phase 0.5 spectra. The result is two apparent cyclotron humps
that peak near 1.58 and 1.93 $\mu$m.}
\end{figure}
\end{document}